# A Polynomial-time Nash Equilibrium Algorithm for Repeated Stochastic Games

**Enrique Munoz de Cote**[*]
DEI, Politecnico di Milano
piazza Leonardo da Vinci, 32
20133 Milan, Italy
munoz@elet.polimi.it

**Michael L. Littman**[†]
Dept. of Computer Science
Rutgers University
Piscataway, NJ 08854
mlittman@cs.rutgers.edu

## Abstract

We present a polynomial-time algorithm that always finds an (approximate) Nash equilibrium for repeated two-player stochastic games. The algorithm exploits the folk theorem to derive a strategy profile that forms an equilibrium by buttressing mutually beneficial behavior with threats, where possible. One component of our algorithm efficiently searches for an approximation of the egalitarian point, the fairest pareto-efficient solution. The paper concludes by applying the algorithm to a set of grid games to illustrate typical solutions the algorithm finds. These solutions compare very favorably to those found by competing algorithms, resulting in strategies with higher social welfare, as well as guaranteed computational efficiency.

## 1  Problem Statement

Stochastic games (Shapley, 1953) are a popular model of multiagent sequential decision making in the machine-learning community (Littman, 1994; Bowling & Veloso, 2001). In the learning setting, these games are often repeated over multiple rounds to allow learning agents a chance to discover beneficial strategies.

Mathematically, a two-player stochastic game is a tuple $\langle \mathcal{S}, s_0, A_1, A_2, \mathcal{T}, U_1, U_2, \gamma \rangle$; namely, the set of states $\mathcal{S}$, an initial state $s_0 \in \mathcal{S}$, action sets for the two agents $A_1$ and $A_2$, with joint action space $\mathcal{A} = A_1 \times A_2$; the state-transition function, $\mathcal{T} : \mathcal{S} \times \mathcal{A} \to \Pi(\mathcal{S})$ ($\Pi(\cdot)$ is the set of probability distributions over $\mathcal{S}$); the utility functions for the two agents $U_1, U_2 : \mathcal{S} \times \mathcal{A} \to \Re$, and the discount $0 \leq \gamma \leq 1$.

[*] Supported by The National Council of Science and Technology (CONACyT), Mexico, under grant No. 196839.
[†] Supported, in part, by NSF IIS-0325281.

In an infinitely repeated stochastic game, the stochastic game is played an unbounded number of rounds. On each round, a stage game is played, starting in $s_0$ and consisting of a series of state transitions (steps), jointly controlled by the two agents. At each step, both agents simultaneously select their actions, possibly stochastically, via *strategies* $\pi_i$ (for each agent $i$). To avoid infinitely long rounds, after each step, the round is allowed to continue with probability $\gamma$, otherwise it is terminated. The payoff for a player in a stage game is the total utility obtained before the stage game is terminated. (Note that the continuation probability $\gamma$ is equivalent to a discount factor.) Players behave so as to maximize their average stage-game payoffs over the infinite number of rounds.

A strategy profile, $\pi = \langle \pi_1, \pi_2 \rangle$, is a *Nash equilibrium* (NE) if each strategy is optimized with respect to the other. In an equilibrium, no agent can do better by changing strategies given that the other agent continues to follow its strategy in the equilibrium. In a repeated game, the construction of equilibrium strategy profiles can involve each player changing strategy from round to round in response to the behavior of the other agent. Note that an $\epsilon$-approximate NE is one in which no agent can do better by more than $\epsilon$ by changing strategies given that the other agent continues to follow its strategy in the equilibrium.

Our approach to finding an equilibrium for repeated stochastic games relies on the idea embodied in the *folk theorems* (Osborne & Rubinstein, 1994). The relevant folk theorem states that if an agent's performance is measured via *expected average* payoff, for any *strictly enforceable* (all agents receive a payoff larger than their minimax values) and *feasible* (payoffs can be obtained by adopting some strategy profile) set of average payoffs to the players, there exist equilibrium strategy profiles that achieve these payoffs. The power of this folk theorem is that communally beneficial play, such as mutual cooperation in the Prisoner's Dilemma, can be justified as an equilibrium. A conceptual drawback is that there may exist infinitely many feasible and enforceable payoffs (and therefore a daunting set of equilibrium strategy profiles to choose from). We focus on the search for a special point inside this (possibly infinite) set of solutions that maximizes the minimum advantage obtained by the players. (The *advantage* is the improvement a player gets over the payoff it can guarantee by playing defensively.) We call this point the egalitarian point, after Greenwald and Hall (2003). Other points can also be justified, such as the one that maximizes the product of advantages—the Nash bargaining solution (Nash, 1950).

Earlier work (Littman & Stone, 2005) has shown that the folk theorem can be interpreted computationally, resulting in a polynomial-time algorithm for repeated games. In the prior work, the game in each round is represented in matrix form—each strategy for each player is explicitly enumerated in the input representation. This paper considers the analogous problem when each stage game is represented much more compactly as a stochastic game. Representing such games in matrix form would require an infinitely large matrix since the number steps per round, and therefore the complexity of the strategies, is unbounded. Even if we limit ourselves to stationary deterministic strategies, there are exponentially many to consider.

Concretely, we address the following computational problem. Given a stochastic game, return a strategy profile that is a Nash equilibrium—one whose payoffs match those of the egalitarian point—of the average payoff repeated stochastic game in polynomial time. In fact, because exact Nash equilibria in stochastic games can require unbounded precision, our algorithm returns an arbitrarily accurate approximation.

## 2 Background

Here, we present background on the problem.

### 2.1 Minimax Strategies

Minimax strategies guarantee a minimum payoff value, called the security value, that an agent can guarantee itself by playing a *defensive* strategy. In addition, an agent can be held to this level of payoff if the other agent adopts an aggressive *attack* strategy (because minimax equals maximin). Given that minimax strategies guarantee a minimum payoff value, no rational player will agree on any strategy in which it obtains a payoff lower than its security value. The pair of security values in a two-player game is called the *disagreement point*.

The set $X \subseteq \mathbb{R}^2$ of average payoffs achievable by strategy profiles can be visualized as a region in the x-y plane. This region is convex because any two strategy profiles can be mixed by alternating over successive rounds to achieve joint payoffs that are any convex combination of the joint payoffs of the original strategy profiles. The disagreement point $v = (v_1, v_2)$ divides the plane into two regions (see Figure 1): a) the region of mutual advantages (all points in $X$, above and to the right of $v$), denotes the strictly enforceable payoff profiles; and b) the relative complement of the region of mutual advantage, which are the payoff profiles that a rational player would reject.

In general-sum bimatrix games, the disagreement point can be computed exactly by solving two zero-sum games (von Neumann & Morgenstern, 1947) to find the attack and defensive strategies and their values. In contrast, the solution to any zero-sum stochastic game can be approximated to any degree of accuracy $\epsilon > 0$ via value iteration (Shapley, 1953). The running time is polynomial in $1/(1-\gamma)$, $1/\epsilon$, and the magnitude of the largest utility $U_{\max}$.

### 2.2 Markov Decision Processes

In this paper, we use Markov decision processes (Puterman, 1994), or MDPs, as a mathematical framework for modeling the problem of the two players working together as a kind of *meta*-player to maximize a weighted combination of their payoffs. For any weight $[w, 1-w]$ ($0 \leq w \leq 1$) and point $p = (p_1, p_2)$, define $\sigma_w(p) = wp_1 + (1-w)p_2$.

Note that any strategy profile $\pi$ for a stochastic game has a value for the two players that can be represented as a point $p^\pi \in X$. To find the strategy profile $\pi$ for a stochastic game that maximizes $\sigma_w(p^\pi)$, we can solve MDP($w$), which is the MDP derived from replacing the utility $r = (r_1, r_2)$ in each state with $\sigma_w(r)$.

### 2.3 Other Solutions for Stochastic Games

There are several solution concepts that have been considered in the literature. Generally speaking, a Nash equilibrium (NE) is a vector of independent strategies in which all players optimize their independent probability distributions over actions with respect to expected payoff. A correlated equilibrium (CE) allows for dependencies in the agent's randomizations, so a CE is a probability distribution over *joint* spaces of actions. Minimax strategies maximize payoff in the face of their worst opponent. At the other extreme, "friend" strategies maximize behavior assuming the opponents are working to maximize the agent's own utility. Friend strategies are appropriate in purely cooperative settings but can perform very badly in mixed incentive settings.

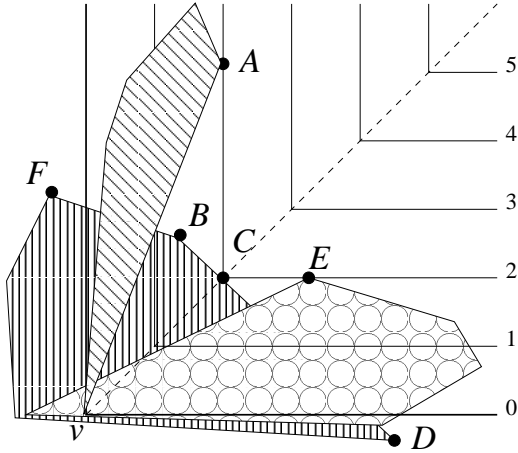

Figure 1: The convex hull $X \subseteq \mathbb{R}^2$ of all payoff profiles. Point $v \in X$ depicts the disagreement point. Three different convex hulls are illustrated. The diagonal line is the egalitarian line.

A powerful and general algorithmic approach for sequential decision problems is value iteration (Bellman, 1957). Variants of value iteration can be defined for each of the solution concepts described above (Zinkevich et al., 2005).

## 3 Algorithm Description

Of all feasible Nash equilibria, we are interested in one whose payoffs match the egalitarian point, thus maximizing the minimum of the payoffs of the two players. Mathematically, we are searching for a point $P = \mathrm{argmax}_{x \in X} \min_v(x)$. Here, $\min_v(x)$ is the *egalitarian value* of $x$, meaning $\min(x_1 - v_1, x_2 - v_2)$ where $x = (x_1, x_2)$ and $v$ is the disagreement point. Note that $\min_v(P) \geq 0$ because $X$ is convex and $v$ is the disagreement point—there is a strategy in which both players do at least as well as $v$.

Define the *egalitarian line* to be the line corresponding to the payoffs in which both player's payoffs are equally high above the disagreement point. Consider the two "friend" solutions to the game, where $L$ is the value to the two players when maximizing Player 2's payoff and $R$ is the value to the two players when maximizing Player 1's payoff. Because $X$ is convex, the egalitarian point $P$ is either $L$, $R$, or the (highest) intersection between $X$ and the egalitarian line.

Figure 1 illustrates these three situations with three different example $X$ sets. The solid L-shaped lines in the figure are the contour lines with equal egalitarian values. The egalitarian point in $X$ is the one that reaches the topmost contour line. In the set filled with diagonal lines, the point $A$ is the egalitarian point and $\min_v(A) = 2$. All other points are to the left of $A$, so $A$ is the point with maximum x coordinate. In the set filled with circles, $E$ is the egalitarian point with $\min_v(E) = 2$. All other points are below $E$, so $E$ is the point with maximum y coordinate.

The intermediate region with the vertical fill lines is a bit more complex. Point $F$ has the largest y coordinate, but its egalitarian value is negative because of the x coordinate. Point $D$ has the largest x coordinate, but its egalitarian value is negative because of the y coordinate. The point $C$, which is a linear combination of the vertices $B$ and $D$, is the egalitarian point, again with a value of 2.

Finding points like $E$ and $A$ is easy—it is just a matter of solving the "friend" MDPs derived using weights of $[0, 1]$ and $[1, 0]$ and halting if either $R$ is on the left or $L$ is on the right of the egalitarian line. However, finding point $C$ is harder, since we need to find points $B$ and $D$ and then intersect the line between them with the egalitarian line.

Figure 2 presents the overall FolkEgal algorithm for finding a strategy profile that achieves the egalitarian point in stochastic games. FolkEgal$(U_1, U_2, \epsilon)$ works for utility functions $U_1$ and $U_2$ and seeks an equilibrium with accuracy $\epsilon$. The routine $(\delta_i, \alpha_{-i}, v_i) := \mathrm{Game}(U_i, \epsilon)$ solves the zero-sum game with utility function $U_i$ to accuracy $\epsilon$. It returns $v_i$ and $\delta_i$, which are the value and strategy (respectively) for the maximizing player $i$, and $\alpha_{-i}$, which is the attack strategy of $i$'s opponent $(-i)$. We do not provide code for this subroutine as any zero-sum stochastic game solver can be used. Littman and Stone (2005) provide details on using the attack strategies to stabilize the discovered mutually beneficial payoffs, which we do not repeat here.

The key missing subroutine is $(P, \pi) := \mathrm{EgalSearch}(L, R, T)$, which finds the intersection between the convex hull of payoffs with the egalitarian line. It returns the egalitarian point $P$ and a strategy profile $\pi$ that achieves it. It is given a point $L \in X$ to the left of the egalitarian line, a point $R \in X$ to the right of the egalitarian line, and a bound $T$ on the number of iterations. Section 4 explains how to choose $T$.

The algorithm is laid out in Figure 3. Its basic structure is a kind of binary search. On each iteration, it solves an MDP to try to find a policy closer to the egalitarian line. It makes use of several support subroutines. The call $w := \mathrm{Balance}(L, R)$ returns the weight $w$ for which $\sigma_w(L) = \sigma_w(R)$. It can be found by solving a linear equation. The payoff of the optimal strategy profile $\pi$ for $w$ should be an improvement on $L$ and $R$ with respect to the weight $w$, that is,

```
Define FolkEgal(U_1, U_2, ε):
    // Find "minimax" strategies
    Let (δ_1, α_2, v_1) := Game(U_1, ε/2)
    Let (δ_2, α_1, v_2) := Game(U_2, ε/2)
    // Make the disagreement point the origin
    Let v := (v_1, v_2)
    Let U_1 := U_1 - v
    Let U_2 := U_2 - v
    // Find "friend" strategies
    Let (R_0, π_2) := MDP([1, 0])
    Let (L_0, π_1) := MDP([0, 1])
    // Find the egalitarian point and its policy
    If R is left of the egalitarian line:
        Let (P, π) := (R_0, π_2)
    Elseif L is to the right of the egalitarian line:
        Let (P, π) := (L_0, π_1)
    Else:
        Let (P, π) := EgalSearch(L_0, R_0, T)
    // If game is like zero sum, compete
    If min_v(P) ≤ ε:
        Return (δ_1, δ_2)
    // Else, mutual advantage
    Return π, modified to use threat strategies
        α_1 and α_2 to enforce the equilibrium
```

Figure 2: Our approach to finding the egalitarian point and a strategy profile that achieves it.

```
Define EgalSearch(L, R, T):
    If T = 0:
        Return Intersect(L, R)
    Let w := Balance(L, R)
    Let (P, π) := MDP(w)
    If P · w = L · w:
        Return Intersect(L, R)
    If P is to the left of egalitarian line:
        Return EgalSearch(P, R, T - 1)
    Else:
        Return EgalSearch(L, P, T - 1)
```

Figure 3: Our search algorithm for intersecting the convex payoff region with the egalitarian line.

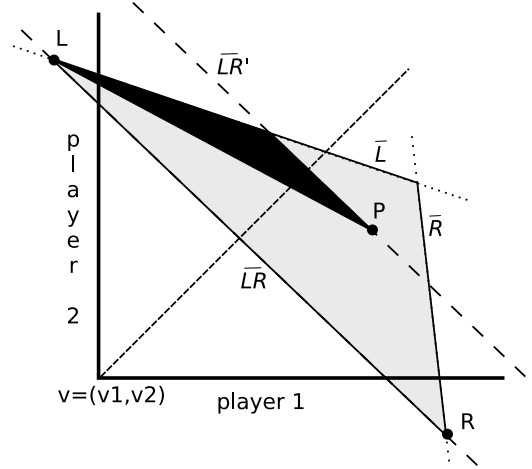

Figure 4: An illustration of the behavior of EgalSearch.

$\sigma_w(P^\pi) \geq \sigma_w(R) = \sigma_w(L)$. If it is not strictly better, the search ends. Otherwise, the new point is used as either $L$ or $R$ and it continues.

The final strategy profile returned is found via Intersect($L, R$), which discovers the right way to alternate between $L$ and $R$ to produces a payoff on the egalitarian line. Again, a simple linear equation suffices to identify this strategy profile.

Figure 4 illustrates a step of the algorithm. First, note the disagreement point $v$ and the egalitarian line heading out from it. The algorithm is given points $L$ and $R$ such that $L$ is on the left of the egalitarian line and $R$ is on the right. In the diagram, the line passing through $L$ labeled $\bar{L}$ is the set of points $p$ such that $\sigma_{w_L}(p) = \sigma_{w_L}(L)$. Since $L$ was returned as the maximum payoff with respect to some weight $w_L$, none of the points in the convex set can be above this line. Similarly, $w_R$ is the weight that was used in the derivation of $R$ and therefore no payoffs are possible beyond the $\bar{R}$ line in the figure.

Next, notice that both $L$ and $R$ are the payoffs for some strategy profile, so both lie in the convex set. Furthermore, any payoff on the line between $L$ and $R$ can also be achieved by some strategy profile. The weight $w$, derived by Balance($L, R$), is the weight such that every payoff $p$ along the line between $L$ and $R$ has the same weighted value $\sigma_w(p)$. The line is called $\bar{LR}$ in the figure.

Putting these ideas together, consider what happens when we solve MDP($w$). We know the result will be at least as high as the $\bar{LR}$ line, since we already know there is a strategy profile that can achieve this payoff. However, we also know that it can't go above the $\bar{R}$ and $\bar{L}$ lines. So, the solution is constrained to lie inside the gray shaded triangle in the figure.

The point $P$ is the hypothetical solution to MDP($w$). Since it is on the right of the egalitarian line, it replaces $R$ in the next iteration. The black triangle represents the region that will be searched in the next iteration. In the next section, we show that each iteration reduces the area of this triangle substantially, and thus that a small number of iterations are needed to reduce its intersection with the egalitarian line to $ε/2$.

## 4 Algorithm Analysis

The main open issue is to set the parameter $T$, which controls the maximum number of search iterations in EgalSearch. Since solving MDPs and the various other steps each take polynomial time, the overall runtime of FolkEgal is polynomial if and only if $T$ is bounded by a polynomial.

Let's say we are given a triangle with area $\nu$ where the corners are possible joint payoffs. Let point $p$ be the point in the triangle that maximizes $\min_v(p)$. Let point $r$ be the point along the longest edge of the triangle that maximizes $\min_v(r)$.

**Claim 1**: $\min_v(p) - \min_v(r) \leq \sqrt{2\nu}$.

To see why, let's consider two facts.

1. For points $x$ and $y$, if $\|x - y\|_2 \leq \delta$, then $y = x + \Delta$ for some $\Delta = (\Delta_1, \Delta_2)$ where $|\Delta_1| \leq \delta$ and $|\Delta_2| \leq \delta$. It follows that $\min_v(y) = \min_v(x+\Delta) \leq \min_v(x + \delta) = \min_v(x) + \delta$. Reversing $x$ and $y$, we find $|\min_v(x) - \min_v(y)| \leq \delta$.

2. A triangle with longest edge $b$ must have an altitude, $h$, where $h \leq b$, otherwise it would not fit inside the triangle. Therefore, its area is $\nu = 1/2bh \geq 1/2h^2$. Thus, $h \leq \sqrt{2\nu}$. This argument shows that for any point $x$ in the triangle and $y$ on the largest side, $\|x - y\|_2 \leq h \leq \sqrt{2\nu}$.

Combining these two facts proves the claim: $|\min_v(p) - \min_v(r)| \leq \sqrt{2\nu}$.

**Claim 2**: Figure 4 shows the generic situation in which the algorithm has found points $L$ and $R$ using weights that result in the edges labeled with $\bar{L}$ and $\bar{R}$. The gray triangle is the remaining region to search. The algorithm then performs an optimization that uncovers a point $P$ inside this region using weight $w$. The process then repeats with the black triangle. Note:

1. The angle at the "top" of the triangle gets wider each iteration.

    The fact follows because the new top vertex is interior to the main triangle.

2. The area of the black triangle is less than or equal to half of that of gray triangle.

    You can visualize the gray triangle as consisting of three shapes—a top triangle, a trapezoid, and the black triangle. Note that the black triangle and the trapezoid share the same height, but the large base of the trapezoid ($\bar{L}R$ line) is longer than the base of the black triangle on line $\bar{L}R'$ (because the gray triangle tapers). Therefore, the black triangle is smaller than the trapezoid and so is less than half of the area of the gray triangle.

Combining the claims, let $p_T$ be the point that maximizes $\min_v(p_T)$ in the triangle active in the $T$th iteration. Let $r_T$ be the point that maximizes $\min_v(r_T)$ on the longest edge of the triangle active in the $T$th iteration. Let $\nu_T$ be the area of the triangle active in the $T$th iteration.

1. $\nu_T \leq \nu_0 \times 1/2^T$

2. $\min_v(p_T) \leq \min_v(r_T) + \sqrt{2\nu_T}$.

So, $\min_v(p_T) \leq \min_v(r_T) + \sqrt{2\nu_0/2^T}$.

If we want the difference to be smaller than, say, $\epsilon$, we need $\sqrt{2\nu_0/2^T} \leq \epsilon$, or $T \geq \log(2\nu_0/\epsilon^2)$.

Since $\nu_0 \leq U_{\max}^2$, the number of iterations is polynomially bounded in the main parameters of the problem and the approximation factor. Each iteration runs in polynomial time.

## 5 Experimental Results

To illustrate the kinds of equilibria produced by our algorithm and to compare them to existing algorithmic approaches, we devised a family of stochastic games played on grids.

All are games played by players A and B. The grids differ in structure, but they all use the same dynamics. Both players A and B occupy distinct cells of the grid and can choose one of 5 distinct actions: N, S, E, W and stand. Actions are executed simultaneously and transitions from one cell to another are deterministic unless a) there is a semi-passable wall in between cells (depicted as a dotted wall in Figure 5(b)), in which case the player transitions to the desired cell with probability 1/2, or, b) both players attempt to step into the same cell, where the collision is resolved at random by a coin flip, so only one player ends up occupying the desired cell and the other makes no transition. Walls are impassable and players cannot pass through each other—attempts to do so result in no transition.

Goal locations can be specific for some agent X (depicted as a dollar sign with subindex, for example, 5(a), 5(c), 5(d), 5(e)) or general (depicted as a dollar sign without subindex, for example Figures 5(b) and 5(c)). The game ends after any player gets to one of its goals and a goal reward is given. There is a step cost for each of the actions N, S, E, W, but stand has no cost. Note that games are general sum in that it is possible for both players to score by both reaching goals simultaneously.

All games use $\gamma = 0.95$, \$ = \$$_A$ = \$$_B$ = 100 and step cost = $-1$. One exception is that the step cost = $-10$ for the asymmetric game[1].

---

[1]This example can be reconstructed with a step cost of

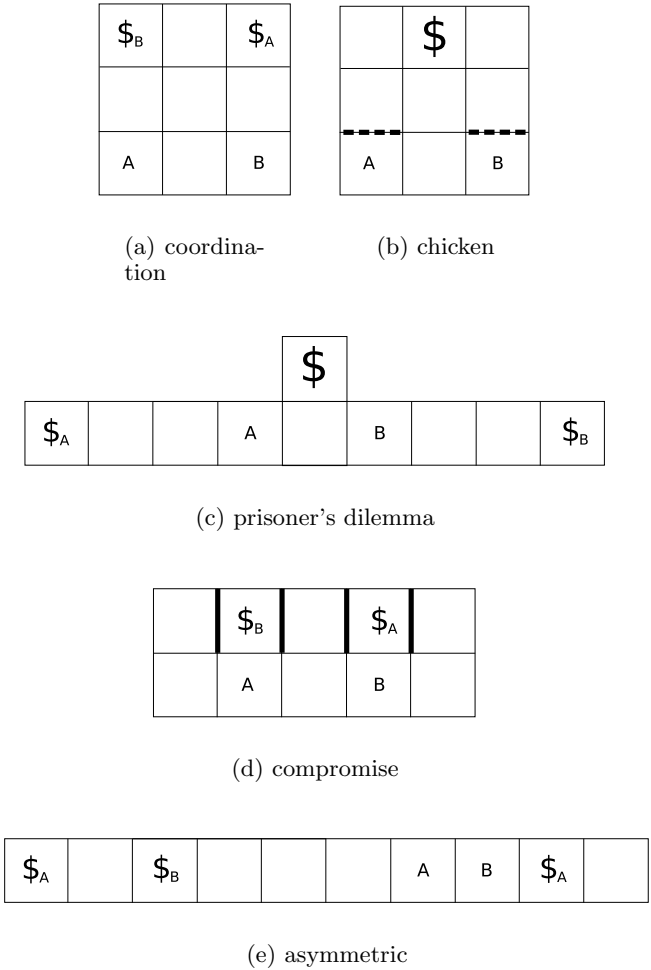

Figure 5: Grid games in their initial state. $\$_X$ = goal for agent X, $\$$ = common goal

(a) coordination
(b) chicken
(c) prisoner's dilemma
(d) compromise
(e) asymmetric

In our results, we include runs for four solution algorithms. Security-VI uses minimax for each player. It is guaranteed to find an equilibrium in zero-sum games, but not in general. Friend-VI uses a self-regarding optimization strategy for both players—each behaves under the assumption that the other player is trying to help it (Littman, 2001). Such an approach can work well in identical payoff games, but since policies are computed independently, it need not. CE-VI[2] computes a correlated equilibrium for the players at each state. As a result, actions are guaranteed to be coordinated, but such an algorithm need not converge to a Nash equilibrium in general. Our FolkEgal algorithm will always find a Nash equilibrium of the repeated game.

---

one, but many more states are needed, so we decided to keep the example small by modifying the step cost.

[2]CE-VI stands for all variants (Greenwald & Hall, 2003) of CE (uCE, eCE, rCE), as their results were identical.

We now present results with the set of grid games in Figure 5. For each game, for each solution algorithm, we present the expected payoffs for each agent along with an informal description of the returned strategy profiles.

### 5.1 Coordination

| algorithm | agent 1 | agent 2 |
|---|---|---|
| security-VI | 0.0 | 0.0 |
| friend-VI | 45.7 | 45.7 |
| CE-VI | 82.8 | 82.8 |
| FolkEgal | 82.8 | 82.8 |

This game is not terribly interesting, but the fact that the players need to pass by each other without colliding makes it relevant to consider their interaction. Friend-VI is unable to coordinate with the opponent and sometimes will collide. Security-VI finds that the worst opponent can always block the goal, so players stay still forever to avoid step costs. Both CE-VI and FolkEgal find a Nash equilibrium by avoiding each other *en route* to goal, and both achieve optimal behavior.

### 5.2 Chicken

| algorithm | agent 1 | agent 2 |
|---|---|---|
| security-VI | 43.7 | 43.7 |
| friend-VI | 42.7 | 42.7 |
| CE-VI | 88.3 | 43.7 |
| FolkEgal | 83.6 | 83.6 |

This game has an element of the game "chicken" in that both players prefer taking the center path, but given that the other player is taking the center path, the side path is more attractive. We used a variation of standard grid game (Hu & Wellman, 2003) in which collisions are resolved by a coin flip (Littman, 1994) and there is no explicit collision cost.

The difference between security-VI and friend-VI in this game is how an agent behaves if it cannot make it to the center square. In the defensive strategy, once a player does not get the center it will stay put because it assumes (rightly) that the opponent will proceed directly to the goal. Friend-VI, on the other hand, will naively continue moving toward the goal under the assumption that the other player will let it pass. It incurs a small step cost for its overly optimistic outlook.

CE-VI finds an asymmetric solution in which one player is assigned to take the center and the other one uses the side passage, through the semi-passable wall. This policy is a Nash equilibrium in that neither player can improve its reward unilaterally.

FolkEgal finds the solution halfway (0.5) along the edge between vertices $(83.14, 84.05)$ and $(84.05, 83.14)$. These points correspond to strategies in which one player takes the center and continues beside the goal, waiting for the other player to catch up. The two players reach the goal at the same time. The first player to go through the center incurs a slightly higher cost because it must step around the goal before waiting, hence the asymmetric (but close) values. The weight of .5 means that each strategy is played with equal frequency (strict alternation, say), in the equilibrium.

Note that both players score slightly worse than the dominant player found in the CE-VI solution, due to the cost of coordination. However, both the value obtained by the minimum player *and* the total reward for the two players is better for the FolkEgal algorithm than for CE-VI.

### 5.3 Prisoner's Dilemma

| algorithm | agent 1 | agent 2 |
|---|---|---|
| security-VI | 46.5 | 46.5 |
| friend-VI | 46.0 | 46.0 |
| CE-VI | 46.5 | 46.5 |
| FolkEgal | 88.8 | 88.8 |

This game was designed to mimic the Prisoner's dilemma. The main choice faced by each of the two agents is whether to move toward the shared goal location in the center or whether to attempt to reach the goal location further out on the side. If both players move toward the center, each has a 50–50 chance of making it to the goal in two steps. If both players move toward the sides, each has a 100% chance of reaching the goal in 3 steps. Clearly, moving to the side is better. However, whichever decision its opponent makes (side or center), the player scores higher by moving to the center.

The results closely match what happens when bimatrix-game versions of the algorithms are applied to Prisoner's dilemma. Security-VI and CE-VI find strategies where both players move to the center (defect). This strategy profile is a Nash equilibrium. Friend-VI is similar, although (as above) once a player is unable to take the center, it continues to try to do so assuming the other player will voluntarily get out of the way. Again, this behavior results in additional unnecessary step costs and is not an equilibrium.

We had expected FolkEgal to find a solution where both players move to their side goals, reserving the center square as a threat (much like tit-for-tat). In fact, FolkEgal found a slightly better scheme—one agent gets the closer common goal (saving step costs) but waits for the other to get to its private goal before entering. FolkEgal find points $(89.3, 88.3)$ and $(88.3, 89.3)$, and players alternate.

### 5.4 Compromise

| algorithm | agent 1 | agent 2 |
|---|---|---|
| security-VI | 0.0 | 0.0 |
| friend-VI | $-20.0$ | $-20.0$ |
| CE-VI | 68.2 | 70.1 |
| FolkEgal | 78.7 | 78.7 |

This game is much like the coordination game, but with the twist that it is not possible for a player to reach its goal without the other player stepping aside.

Security-VI adopts the worst-case assumption that the other player will not step aside. Both players end up staying still, as a result, to avoid step costs. Friend-VI is actually even worse. Since both player assumes the other will step aside, the two players simply ram each other indefinitely.

CE-VI converges to a very interesting strategy. Player A steps into Player B's goal and waits. Player A is blocking Player B from scoring, but it is also allowing Player B to pass. Player B walks to the upper left corner and Player A moves back to its initial position. Note that both players are now 3 steps from their respective goals. At this point, both players move to their goals and arrive simultaneously. One of the more interesting aspects of this strategy profile is that Player B waits in the corner until Player A has stepped out of the goal. The reason is that Player A will not attempt to reach its own goal until it is sure it won't be beaten by Player B. Player B, by keeping a respectful distance, signals to Player A that it is safe to move and both players benefit. Since both players are also choosing actions in their own best interest, the resulting strategy profile is an equilibrium.

This strategy profile, while ingenious (and unexpected to us), does not maximize the value of the minimum player. FolkEgal's solution is to alternate between $L = (79.6, 77.7)$ and $R = (77.7, 79.6)$. The strategy profile, in this case, corresponds to one player moving to the space between the goals, the other moving in front of its goal and waiting, then both players reaching their goals together in 5 steps.

### 5.5 Asymmetric

| algorithm | agent 1 | agent 2 |
|---|---|---|
| security-VI | 0.0 | 0.0 |
| friend-VI | $-200.0$ | $-200.0$ |
| CE-VI | 32.1 | 42.1 |
| FolkEgal | 37.2 | 37.2 |

This game was designed to show how the algorithms react to an asymmetric starting position. Once again, overly optimistic (friend-VI) and pessimistic (security-VI) assumptions result in very low scores for both players.

Note that the players again need to compromise. Without Player B's cooperation, Player A cannot reach its near goal on the right. It also cannot reach its far goal on the left, because Player B can trail behind it and reach its goal before Player A arrives.

CE-VI discovers that Player B can "offer" Player A a compromise by hanging back exactly one square when Player A moves to the left. As a result, both players reach their goal locations on the left simultaneously. The solution is a Nash equilibrium, although it is not the egalitarian solution.

FolkEgal finds the egalitarian solution as a weighted combination of the points $L = (32.13, 42.13)$ and $R = (85, -10)$ with weight approximately .1. Note that point $L$ corresponds to the solution found by CE-VI. Point $R$ corresponds to the strategy where Player B moves to the right and lets Player A reach the near goal location.

### 5.6 Summary

There are a few interesting generalizations to make, based on these results. First, although CE-VI is not guaranteed to find a Nash equilibrium, it did so in all 5 games (including games that were designed specifically to thwart it). We were surprised at the robustness of the algorithm.

CE-VI only found the egalitarian solution in one game, however, whereas FolkEgal found it every time. FolkEgal also has guaranteed polynomial runtime bounds, whereas CE-VI is known to fail to converge in some games (Zinkevich et al., 2005). Friend-VI performed uniformly badly and security-VI, or foe-VI (Littman, 2001), often returned a Nash equilibrium, but one with worse overall performance.

## 6 Future Work

This paper shows how an egalitarian Nash equilibrium solution can be found efficiently for repeated stochastic games. Future work will attempt to generalize these techniques to repeated games on trees (Littman et al., 2006) or DAGs and perhaps even repeated partial information games (Koller et al., 1996).